\documentclass[english]{sbrt}
\usepackage[english]{babel}
\usepackage[utf8]{inputenc}
\usepackage{amsmath}
\usepackage{nccmath}
\usepackage{xcolor}
\usepackage{amssymb}
\usepackage{graphicx}
\usepackage{float}
\usepackage{bm}
\usepackage{xcolor}
\usepackage[justification=centering]{caption}
\usepackage{algorithm}
\usepackage{algorithmic}
\makeatletter
\g@addto@macro\normalsize{%
  \setlength\abovedisplayskip{3pt plus 1pt minus 1pt}%
  \setlength\belowdisplayskip{3pt plus 1pt minus 1pt}%
  \setlength\abovedisplayshortskip{2pt plus 1pt minus 1pt}%
  \setlength\belowdisplayshortskip{2pt plus 1pt minus 1pt}%
}
\makeatother

\begin{document}

\title{Low-Complexity Channel Parameter Estimation for Coprime HRIS-Assisted MIMO Systems}

\author{Jean C. S. Ferreira, Melryllin G. O. Sousa, and Fazal-E Asim
\thanks{\fontsize{6.2}{7.5}\selectfont Jean C. S. Ferreira, Department of Teleinformatics Engineering, Federal University of Ceará, Fortaleza-CE, e-mail: jeancarlos8249@alu.ufc.br; Melryllin G. O. Sousa, Department of Teleinformatics Engineering, Federal University of Ceará, Fortaleza-CE, e-mail: mryllin@alu.ufc.br; Fazal-E Asim, Department of Teleinformatics Engineering, Federal University of Ceará, Fortaleza-CE, e-mail: fazalasim@ufc.br. This work is partially supported by the National Institute of Science and Technology (INCT-Signals) sponsored by Brazil’s National Council for Scientific and Technological Development (CNPq) (Proc. 406517/2022-3), and FUNCAP (Proc. INCT-25255-82587.32.41/64).}}


\maketitle

\markboth{XLIV BRAZILIAN SYMPOSIUM ON TELECOMMUNICATIONS AND SIGNAL PROCESSING - SBrT 2026, SEPTEMBER 29TH TO OCTOBER 2ND, 2026, SALVADOR, BA}{}

\fontsize{9.55}{10.25}\selectfont
\setlength{\parskip}{0pt}

\begin{abstract}
This paper proposes a parameter estimation scheme for uplink MIMO systems assisted by a Hybrid Reconfigurable Intelligent Surface (HRIS). To reduce hardware complexity, the HRIS employs a small number of active elements arranged in a sparse coprime geometry for local sensing, while the remaining elements passively reflect signals. This simultaneous sensing and reflection significantly improves accuracy over fully passive architectures. The method leverages Khatri-Rao and Kronecker factorizations to efficiently decouple the cascaded channel at the base station. Furthermore, spatial smoothing resolves the coprime array rank deficiency, enabling robust angular extraction via Root-MUSIC. Simulations demonstrate highly resilient estimation performance across scenarios.
\end{abstract}

\section{Introduction}

Reconfigurable Intelligent Surfaces (RISs) have emerged as a promising technology for beyond-5G and 6G wireless systems because they enable programmable control of the propagation environment through a large number of low-power reflecting elements. By properly adjusting the phase response of the surface, RIS-assisted systems can improve coverage, suppress interference, and enhance energy efficiency. Nevertheless, these potential gains depend critically on accurate channel state information (CSI). Since conventional passive RISs do not observe the impinging signal directly, the receiver usually has access only to the cascaded UE-RIS-BS channel. This indirect observation makes channel acquisition difficult, especially when the RIS contains a large number of elements and the pilot overhead increases with the channel dimension \cite{ref7,ref2}.

Several approaches have been proposed to reduce the channel estimation burden in RIS-assisted MIMO systems. Tensor and matrix factorization methods exploit the structure of the cascaded channel to separate the UE-RIS and RIS-BS links with reduced training overhead \cite{refLSKRF,refTHz,ref1}. However, in fully passive RIS architectures, the absence of sensing capability at the surface limits parameter identifiability and makes the estimation accuracy highly dependent on the quality of the reflected link. Hybrid RIS (HRIS) architectures address this limitation by equipping part of the surface with active sensing elements, thereby providing local measurements of the incident signal while preserving most of the aperture for passive reflection \cite{ref3,ref6}. Compared with designs in which the whole panel is connected to RF chains or partitioned into active subarrays, architectures with only a few active elements are more attractive from the viewpoint of hardware cost, power consumption, and implementation complexity.

In parallel, sparse array processing provides a useful tool for extracting angular information with a limited number of sensors. Coprime arrays, in particular, can generate large virtual apertures and increased degrees of freedom from a small set of physical sensors \cite{ref8,ref5}. This property is well suited to HRIS-aided systems, where only a few elements can be made active without sacrificing the passive reflection gain. However, sparse coprime geometries also introduce missing spatial samples and rank-deficient covariance structures. Coarray-domain spatial smoothing and subspace-based methods, such as MUSIC and Root-MUSIC, can alleviate this issue and enable high-resolution direction finding \cite{ref10,ref9}. These observations motivate the integration of sparse coprime sensing with HRIS channel parameter estimation.

Motivated by the above discussion, this paper investigates parameter estimation for uplink MIMO systems assisted by an HRIS whose active elements follow a sparse "L"-shaped coprime geometry. The proposed scheme combines local sensing at the HRIS with cascaded-channel processing at the base station (BS). The local measurements are first used to estimate the UE departure angle and the HRIS arrival angles. Then, Khatri-Rao and Kronecker factorizations are employed to decouple the cascaded channel and recover the angular parameters associated with the reflection link. Finally, spatial smoothing and Root-MUSIC are applied to overcome the rank deficiency caused by the sparse coprime geometry. Simulation results compare the proposed HRIS architecture with a fully passive RIS baseline under different transmit powers, power-splitting factors, and numbers of active elements. The main contributions of this paper are summarized as follows:
\begin{itemize}
    \item We propose a low-complexity HRIS architecture in which only a few selected elements are active and arranged as a sparse, L-shaped coprime sensing array, while the remaining elements continue to operate as passive reflectors;
    \item We develop a channel parameter estimation framework that combines HRIS local sensing with BS-side cascaded-channel processing through Khatri-Rao and Kronecker factorizations;
    \item We employ coarray-domain spatial smoothing and Root-MUSIC to restore rank and extract angular parameters from the sparse coprime sensing geometry;
    \item We show through simulations that the proposed HRIS strategy improves estimation accuracy compared with a fully passive RIS baseline and remains robust under different power-splitting values and active-element configurations.
\end{itemize}

\textbf{\textit{Notations:}} Bold lowercase ($\bm{a}$) and uppercase ($\bm{A}$) letters denote vectors and matrices, respectively. The operators $(\cdot)^T$, $(\cdot)^H$, and $(\cdot)^{-1}$ represent the transpose, Hermitian transpose, and matrix inverse, respectively, while $\text{vec}(\cdot)$ stacks the columns of a matrix into a vector. The symbols $\otimes$, $\odot$, and $\diamond$ denote the Kronecker, Hadamard, and Khatri-Rao products, respectively. The operator $\text{diag}(\bm{v})$ forms a diagonal matrix from the vector $\bm{v}$. Throughout the paper, steering vectors are written as functions of their spatial frequencies, and the identity $\text{vec}(\bm{A}\text{diag}(\bm{v})\bm{B}) = (\bm{B}^T \diamond \bm{A})\bm{v}$ is used to express and decouple the RIS-assisted cascaded channel.

\section{System Model}
\begin{figure}
    \centering
    \includegraphics[width=0.7\linewidth]{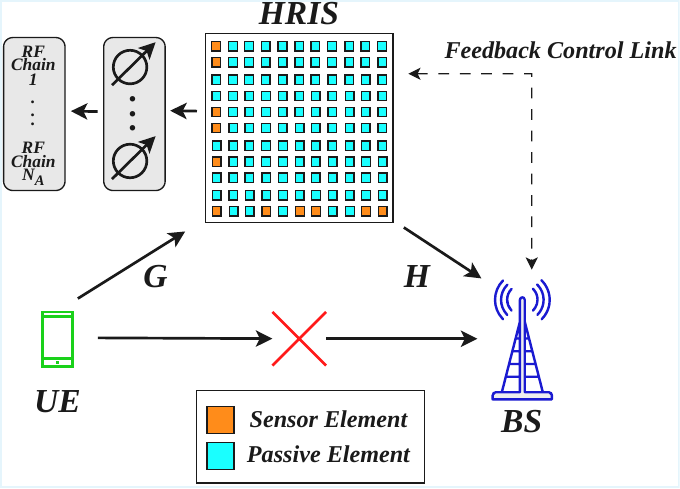}
    \caption{HRIS Assisted MIMO System Model.}
    \label{fig:systemDiagram}
\end{figure}

Consider an uplink MIMO system in which a base station (BS) equipped with $M$ antennas communicates with a user equipment (UE) equipped with $Q$ antennas through an HRIS with $N=N_y \times N_z$ reflecting elements, as illustrated in Fig. \ref{fig:systemDiagram}. Uniform linear arrays (ULAs) are adopted at both the BS and the UE\footnote{\scriptsize Although an uplink scenario with ULA-equipped terminals is considered for clarity, the proposed estimation framework can be extended to downlink transmission and to other two-dimensional array geometries, such as uniform rectangular arrays (URAs) and uniform planar arrays (UPAs).}. The direct UE-BS link is assumed to be blocked, so communication takes place through the UE-HRIS-BS path. The HRIS consists of passive reflecting elements and a small number of active sensing elements. The active elements can simultaneously process part of the incident signal and reflect the remaining part, whereas the passive elements only reflect the impinging signal. Let $\bm{H}$ and $\bm{G}$ denote the HRIS-BS and UE-HRIS channel matrices, respectively \cite{ref1}.

Assuming a dominant line-of-sight (LoS) component, the HRIS-BS channel is modeled as
\begin{equation}
    \bm{H} = \gamma \, \bm{a}(\theta_\text{BS})\bm{b}^T(\phi_{\text{ris}_D}, \theta_{\text{ris}_D}) \in \mathbb{C}^{M \times N},
\end{equation}
where $\gamma$ is the complex path gain of the HRIS-BS link, $\bm{a}(\theta_\text{BS}) \in \mathbb{C}^{M \times 1}$ is the BS receive steering vector, and $\theta_\text{BS}$ is the angle of arrival (AoA) at the BS. The vector $\bm{b}(\phi_{\text{ris}_D},\theta_{\text{ris}_D}) \in \mathbb{C}^{N \times 1}$ represents the HRIS departure steering vector, where $\phi_{\text{ris}_D}$ and $\theta_{\text{ris}_D}$ denote the azimuth and elevation angles of departure from the HRIS, respectively.

Similarly, the UE-HRIS channel is expressed as
\begin{equation}
    \bm{G} = \zeta \, \bm{p}(\phi_{\text{ris}_A}, \theta_{\text{ris}_A}) \bm{q}^T(\phi_\text{UE}) \in \mathbb{C}^{N \times Q},
\end{equation}
where $\zeta$ is the complex path gain of the UE-HRIS link, $\bm{p}(\phi_{\text{ris}_A},\theta_{\text{ris}_A}) \in \mathbb{C}^{N \times 1}$ is the HRIS arrival steering vector, and $\bm{q}(\phi_\text{UE}) \in \mathbb{C}^{Q \times 1}$ is the UE transmit steering vector. The angles $\phi_{\text{ris}_A}$ and $\theta_{\text{ris}_A}$ are the azimuth and elevation angles of arrival at the HRIS, while $\phi_\text{UE}$ is the UE angle of departure (AoD). In the considered LoS scenario, the path gains are normalized as $\gamma=\zeta=1$.

For half-wavelength antenna spacing, i.e., $\Delta=\lambda/2$, the spatial frequencies associated with the BS and UE ULAs are defined as $\mu_\text{BS}=\pi\sin(\theta_\text{BS})$ and $\psi_\text{UE}=\pi\sin(\phi_\text{UE})$, respectively. Hence, the corresponding ULA steering vectors are given by
\begin{equation}
    \bm{a}(\mu_\text{BS}) = \left[1, e^{-j\mu_\text{BS}}, \dots, e^{-j(M-1)\mu_\text{BS}}\right]^T \in \mathbb{C}^{M \times 1},
\end{equation}
\begin{equation}
    \bm{q}(\psi_\text{UE}) = \left[1, e^{-j\psi_\text{UE}}, \dots, e^{-j(Q-1)\psi_\text{UE}}\right]^T \in \mathbb{C}^{Q \times 1}.
\end{equation}

The HRIS elements are arranged as an $N_y \times N_z$ URA on the $y$-$z$ plane, with $N=N_yN_z$. Its two-dimensional response is represented through the Kronecker product of the one-dimensional responses along the horizontal and vertical axes. For the HRIS-BS link, define $\mu_{\text{ris}_D}=\pi\sin(\theta_{\text{ris}_D})\sin(\phi_{\text{ris}_D})$ and $\psi_{\text{ris}_D}=\pi\cos(\theta_{\text{ris}_D})$. The HRIS departure steering vector is then written as
\begin{equation}
    \label{eq:steering_RIS}
    \bm{b}(\mu_{\text{ris}_D}, \psi_{\text{ris}_D}) = \bm{b}_y(\mu_{\text{ris}_D}) \otimes \bm{b}_z(\psi_{\text{ris}_D}) \in \mathbb{C}^{N \times 1},
\end{equation}
where
\begin{equation}
    \bm{b}_y(\mu_{\text{ris}_D}) = \left[1, e^{-j\mu_{\text{ris}_D}}, \dots, e^{-j(N_y-1)\mu_{\text{ris}_D}}\right]^T  \in \mathbb{C}^{N_y \times 1},
\end{equation}
\begin{equation}
    \bm{b}_z(\psi_{\text{ris}_D}) = \left[1, e^{-j\psi_{\text{ris}_D}}, \dots, e^{-j(N_z-1)\psi_{\text{ris}_D}}\right]^T \in \mathbb{C}^{N_z \times 1}.
\end{equation}
The HRIS arrival steering vector $\bm{p}(\mu_{\text{ris}_A},\psi_{\text{ris}_A}) \in \mathbb{C}^{N \times 1}$ is defined analogously by using the spatial frequencies $\mu_{\text{ris}_A}$ and $\psi_{\text{ris}_A}$ associated with the incident UE-HRIS signal. Note that the spatial steering vectors depend solely on element positioning and signal wavelength, remaining physically valid and structurally identical for both passive and coprime active elements.

The active sensing elements are connected to individual radio-frequency (RF) chains and are placed according to a sparse ``L''-shaped coprime geometry \cite{ref2}. Let $P_c$ and $Q_c$ be coprime integers defining the candidate sensor locations along one HRIS axis. The corresponding coprime position set is
$\mathbb{S}_{\text{co}} = \{0,P_c,2P_c,\dots,(Q_c-1)P_c,Q_c,2Q_c,\dots,(2P_c-1)Q_c\}$.
For a finite HRIS aperture, only the entries of $\mathbb{S}_{\text{co}}$ that lie within the physical axis are retained, forming the practical index set $\mathbb{S}$. For instance, $P_c=3$ and $Q_c=5$ on an 11-element axis yield $\mathbb{S}=\{0,3,5,6,9,10\}$. Since the origin is shared by the horizontal and vertical branches of the L-shaped array, the total number of active sensing elements is $N_A=2|\mathbb{S}|-1$, with $N_A\ll N$ \cite{ref3}. This sparse placement provides sensing capability with a small number of RF chains while preserving most HRIS elements for passive reflection.

During the $k$-th transmission block, the signal received at the BS is modelled as
\begin{equation}
\begin{split}
    \bm{Y}^\text{BS}_k &= \bm{H} \text{diag}(\bm{\rho}) \text{diag}(\bm{\omega}_k) \bm{G} \bm{S} + \bm{V}_k \\
    & = \bm{H} \text{diag}(\bm{\rho} \odot \bm{\omega}_k) \bm{G} \bm{S} + \bm{V}_k \in \mathbb{C}^{M \times Q},
\end{split}
\end{equation}
where $\bm{S}\in\mathbb{C}^{Q\times Q}$ is the orthogonal pilot matrix transmitted by the UE, $\bm{\omega}_k\in\mathbb{C}^{N\times 1}$ contains the HRIS reflection coefficients applied during the $k$-th block, and $\bm{V}_k\in\mathbb{C}^{M\times Q}$ is the additive white Gaussian noise (AWGN) at the BS. The vector $\bm{\rho}=[\rho_1,\dots,\rho_N]^T$ controls the reflected power fraction at the HRIS. For an active element indexed by $n\in\mathbb{S}$, $0\le\rho_n\le1$; for a purely passive element, $\rho_n=1$. Thus, $\rho_n$ represents the fraction of incident power reflected by the $n$-th HRIS element toward the BS \cite{ref2}.

For local sensing, define $\bm{\rho}_{\mathbb{S}}=[\rho_1^{\mathbb{S}},\dots,\rho_{N_A}^{\mathbb{S}}]^T$ as the vector collecting the reflection fractions of the active elements. The corresponding sensing factors are given by $\bm{\eta}^{\mathbb{S}}=[\eta_1^{\mathbb{S}},\dots,\eta_{N_A}^{\mathbb{S}}]^T$, where $\eta_i^{\mathbb{S}}=\sqrt{1-\rho_i^{\mathbb{S}}}$. Hence, $\eta_i^{\mathbb{S}}$ determines the portion of the incident signal absorbed and processed by the $i$-th active sensing element. Let $\bm{J}_{\mathbb{S}}\in\{0,1\}^{N_A\times N}$ be the selection matrix that extracts the rows of $\bm{G}$ associated with the active elements. The signal received at the HRIS sensing chains is then
\begin{equation}
\label{eq:Y_sens}
\begin{split}
    \bm{Y}^\text{HRIS} &= \text{diag}(\bm{\eta}^{\mathbb{S}}) \bm{J}_{\mathbb{S}} \bm{G} \bm{S} + \bm{N}^\text{HRIS} \\
    &= \bm{G}^{\mathbb{S}} \bm{S} + \bm{N}^\text{HRIS} \in \mathbb{C}^{N_A \times Q},
\end{split}
\end{equation}
where $\bm{G}^{\mathbb{S}}=\text{diag}(\bm{\eta}^{\mathbb{S}})\bm{J}_{\mathbb{S}}\bm{G}\in\mathbb{C}^{N_A\times Q}$ is the effective UE-HRIS channel observed by the active sensing elements, and $\bm{N}^\text{HRIS}$ denotes the AWGN matrix at the HRIS sensing receiver.

\section{Channel Parameter Estimation Scheme}
\begin{algorithm}[htbp]
\caption{Proposed HRIS Channel Parameter Estimation}
\label{alg:proposed_hris}
\begin{algorithmic}[1]
\REQUIRE $\bm{Y}^{\text{HRIS}}$, $\bm{Y}_k^\text{BS}$ ($k=1,\dots,K$), pilot $\bm{S}$, and training matrix $\bm{\Omega}$.

\STATE \textbf{Phase 1: Local Sensing at HRIS}
\STATE Filter \& Rank-1 extract: $[\hat{\bm{p}}^{\mathbb{S}}, \hat{\bm{q}}^\mathbb{S}] \leftarrow \text{SVD}_1(\bm{Y}^{\text{HRIS}}\bm{S}^H)$
\STATE Extract 1D arrays: $\hat{\bm{p}}^{\mathbb{S}}_y = \bm{J}_y \hat{\bm{p}}^{\mathbb{S}}$, and $\hat{\bm{p}}^{\mathbb{S}}_z = \bm{J}_z \hat{\bm{p}}^{\mathbb{S}}$
\STATE Root-MUSIC: Extract sensing angles $\hat{\phi}_{\text{UE}}^{\mathbb{S}}, \hat{\theta}_{\text{ris}_A}^{\mathbb{S}}, \hat{\phi}_{\text{ris}_A}^{\mathbb{S}}$.

\STATE \textbf{Phase 2: Reflection Processing at BS}
\STATE Filter \& stack blocks: 

$\bm{U} = [\text{vec}(\bm{Y}_1^\text{BS}\bm{S}^H), \dots, \text{vec}(\bm{Y}_K^\text{BS}\bm{S}^H)]$
\STATE Estimate Khatri-Rao channel: $\hat{\bm{E}} = \bm{U}\bm{\Omega}^{\dagger}$
\STATE Factorize: $\{\hat{\bm{H}}, \hat{\bm{G}}\} \leftarrow \text{LSKRF}(\hat{\bm{E}})$, then apply LSKronF.
\STATE Root-MUSIC: 

Extract reflection angles $\hat{\theta}_\text{BS}^\text{Ref}, \hat{\theta}_{\text{ris}_D}^\text{Ref}, \hat{\phi}_{\text{ris}_D}^\text{Ref}, \hat{\phi}_{\text{UE}}^{\text{Ref}}, \hat{\theta}_{\text{ris}_A}^{\text{Ref}}, \hat{\phi}_{\text{ris}_A}^{\text{Ref}}$.

\STATE \textbf{Phase 3: Data Fusion}
\STATE Average shared angles: $\hat{\phi}_{\text{UE}} = \frac{1}{2}(\hat{\phi}_{\text{UE}}^{\mathbb{S}} \!+\! \hat{\phi}_{\text{UE}}^{\text{Ref}})$, $\hat{\theta}_{\text{ris}_A} = \frac{1}{2}(\hat{\theta}_{\text{ris}_A}^{\mathbb{S}} \!+\! \hat{\theta}_{\text{ris}_A}^{\text{Ref}})$, $\hat{\phi}_{\text{ris}_A} = \frac{1}{2}(\hat{\phi}_{\text{ris}_A}^{\mathbb{S}} \!+\! \hat{\phi}_{\text{ris}_A}^{\text{Ref}})$.

\ENSURE Final angles $\hat{\theta}_\text{BS}, \hat{\phi}_\text{UE}, \hat{\theta}_{\text{ris}_D}, \hat{\phi}_{\text{ris}_D}, \hat{\theta}_{\text{ris}_A}, \hat{\phi}_{\text{ris}_A}$.
\end{algorithmic}
\end{algorithm}

The proposed estimation scheme exploits two complementary sources of information provided by the HRIS architecture. First, the active elements locally sense the impinging UE signal and provide direct information about the UE-HRIS link. Second, the BS receives the signal reflected by the complete HRIS aperture, which contains the cascaded UE-HRIS-BS channel. Accordingly, the estimation procedure is divided into: (i) local sensing-based estimation at the HRIS and (ii) reflection-based cascaded channel factorization at the BS.

\subsection{Estimation via Sensing in the HRIS}
\label{sec:EstimationSensing}
From (\ref{eq:Y_sens}), the signal observed by the active HRIS elements is given by $\bm{Y}^{\text{HRIS}}=\bm{G}^{\mathbb{S}}\bm{S}+\bm{N}^{\text{HRIS}}$. Since the pilot matrix is assumed to be orthogonal, i.e., $\bm{S}\bm{S}^H=\bm{I}_Q$, a matched-filter estimate of the effective sensing channel is obtained as
\begin{equation}
    \hat{\bm{G}}^{\mathbb{S}} = \bm{Y}^{\text{HRIS}}\bm{S}^H
    = \bm{G}^{\mathbb{S}} + \bm{N}^{\text{HRIS}}\bm{S}^H.
\end{equation}

The superscript $(\cdot)^{\mathbb{S}}$ indicates quantities associated with the active coprime sensing elements.

Under the dominant LoS model, $\hat{\bm{G}}^{\mathbb{S}}\in\mathbb{C}^{N_A\times Q}$ has an approximately rank-one structure given by the outer product between the HRIS arrival response and the UE transmit response. Hence,
\begin{equation}
    \hat{\bm{G}}^{\mathbb{S}} \approx
    \hat{\bm{p}}^{\mathbb{S}}(\mu_{\text{ris}_A},\psi_{\text{ris}_A})
    \hat{\bm{q}}^{\mathbb{S}}(\psi_{\text{UE}})^T,
\end{equation}
where $\hat{\bm{p}}^{\mathbb{S}}(\mu_{\text{ris}_A},\psi_{\text{ris}_A})\in\mathbb{C}^{N_A\times1}$ is the HRIS steering vector restricted to the active elements and $\hat{\bm{q}}^{\mathbb{S}}(\psi_{\text{UE}})\in\mathbb{C}^{Q\times1}$ is the sensing UE steering vector. These vectors can be estimated by solving the rank-one least-squares problem
\begin{equation}
    \left\{ \hat{\bm{p}}^{\mathbb{S}},\hat{\bm{q}}^{\mathbb{S}} \right\}
    = \arg\min_{\bm{p},\bm{q}}
    \left\| \hat{\bm{G}}^{\mathbb{S}} - \bm{p}\bm{q}^{T} \right\|^2_\text{F}.
\end{equation}
This problem can be solved using a least-squares Kronecker factorization (LSKronF) or, equivalently, by extracting the dominant rank-one factors of $\hat{\bm{G}}^{\mathbb{S}}$.

Due to the sparse L-shaped geometry of the active HRIS elements, the $N_A \times 1$ steering vector $\hat{\bm{p}}^{\mathbb{S}}$ lacks a Kronecker product structure. Instead, the 1D spatial signatures for the horizontal and vertical axes are directly decoupled via selection matrices $\bm{J}_y \in \{0,1\}^{N_{A,y} \times N_A}$ and $\bm{J}_z \in \{0,1\}^{N_{A,z} \times N_A}$, which extract the active elements mapped to each respective axis, yielding
\begin{equation}
    \hat{\bm{p}}^{\mathbb{S}}_y = \bm{J}_y \hat{\bm{p}}^{\mathbb{S}} \in \mathbb{C}^{N_{A,y} \times 1}, \quad \text{and} \quad \hat{\bm{p}}^{\mathbb{S}}_z = \bm{J}_z \hat{\bm{p}}^{\mathbb{S}} \in \mathbb{C}^{N_{A,z} \times 1}.
\end{equation}

The resulting factors provide the spatial signatures used to estimate the UE AoD ($\hat{\phi}_{\text{UE}}^{\mathbb{S}}$) and the HRIS AoA ($\hat{\theta}_{\text{ris}_A}^{\mathbb{S}}, \hat{\phi}_{\text{ris}_A}^{\mathbb{S}}$) sensing parameters.



\subsection{Estimation via Reflection in the HRIS}
\label{sec:EstimationReflection}
During the reflection phase, the BS observes the signal reflected by the complete HRIS aperture. For the $k$-th transmission block, matched filtering of the BS received signal in the system model yields
\begin{equation}
    \bm{U}_k = \bm{Y}_k^\text{BS}\bm{S}^H
    = \bm{H}\text{diag}(\bm{\rho}\odot\bm{\omega}_k)\bm{G}
    + \bm{V}'_k \in \mathbb{C}^{M\times Q},
\end{equation}
where $\bm{V}'_k=\bm{V}_k\bm{S}^H$ is the filtered noise matrix. The vector $\bm{\omega}_k$ contains the HRIS reflection coefficients used in the $k$-th block, while $\bm{\rho}$ accounts for the power-splitting coefficients.

Applying the vectorization operator and the identity $\text{vec}(\bm{A}\text{diag}(\bm{x})\bm{B})=(\bm{B}^T\diamond\bm{A})\bm{x}$ gives
\begin{equation}
    \bm{u}_k = \text{vec}(\bm{U}_k)
    = (\bm{G}^T\diamond\bm{H})(\bm{\rho}\odot\bm{\omega}_k)
    + \bm{v}'_k,
\end{equation}
where $\bm{u}_k\in\mathbb{C}^{MQ\times1}$ and $\bm{v}'_k=\text{vec}(\bm{V}'_k)$. Stacking the observations from $K$ reflection blocks leads to
\begin{equation}
    \bm{U} = (\bm{G}^T\diamond\bm{H})\bm{\Omega}+\bm{V}',
\end{equation}
where $\bm{U}=[\bm{u}_1,\dots,\bm{u}_K]\in\mathbb{C}^{MQ\times K}$, $\bm{V}'=[\bm{v}'_1,\dots,\bm{v}'_K]$, and
\begin{equation}
    \bm{\Omega}=[\bm{\rho}\odot\bm{\omega}_1,\dots,\bm{\rho}\odot\bm{\omega}_K]
    =\text{diag}(\bm{\rho})\bm{W} \in \mathbb{C}^{N \times K}.
\end{equation}

Here, $\bm{W}=[\bm{\omega}_1,\dots,\bm{\omega}_K] \in \mathbb{C}^{N \times K}$ is the HRIS training matrix, which can be designed from columns of a discrete Fourier transform (DFT) matrix. If $\bm{\Omega}$ has full row rank, the Khatri-Rao structured channel can be estimated as
\begin{equation}
    \hat{\bm{E}} = \bm{U}\bm{\Omega}^{\dagger}
    = \bm{G}^T\diamond\bm{H}+\bm{V}'',
\end{equation}
where $(\cdot)^{\dagger}$ denotes the Moore-Penrose pseudoinverse. For the square nonsingular case, $\bm{\Omega}^{\dagger}=\bm{\Omega}^{-1}$.

The matrices $\bm{H}$ and $\bm{G}$ are then recovered from $\hat{\bm{E}}$ by exploiting the column-wise Khatri-Rao structure. This can be formulated as
\begin{equation}
    \left\{ \hat{\bm{H}}, \hat{\bm{G}} \right\}
    = \arg\min_{\bm{H},\bm{G}}
    \left\| \hat{\bm{E}} - \bm{G}^T \diamond \bm{H} \right\|^2_\text{F},
\end{equation}
which can be solved using least-squares Khatri-Rao factorization (LSKRF). After obtaining $\hat{\bm{H}}$ and $\hat{\bm{G}}$, a least-squares Kronecker factorization (LSKronF) decouples the multi-dimensional spatial signatures. For instance, the 2D RIS departure vector $\hat{\bm{b}}$ is factored into its horizontal and vertical 1D components by solving
\begin{equation}
    \left\{ \hat{\bm{b}}^\text{Ref}_y, \hat{\bm{b}}^\text{Ref}_z \right\}
    = \arg\min_{\bm{b}_y,\bm{b}_z}
    \left\| \hat{\bm{b}}^\text{Ref} - \bm{b}_y \otimes \bm{b}_z \right\|^2_\text{F}.
\end{equation}

This identical procedure is systematically applied to extract the remaining 1D steering vectors $\hat{\bm{a}}^\text{Ref}(\mu_\text{BS})$, $\hat{\bm{q}}^\text{Ref}(\psi_\text{UE})$, $\hat{\bm{p}}_y^\text{Ref}(\mu_{\text{ris}_A})$, and $\hat{\bm{p}}_z^\text{Ref}(\psi_{\text{ris}_A})$. The superscript $(\cdot)^{\text{Ref}}$ indicates quantities associated with the HRIS reflection elements.

\subsection{Rank Restoration and Angle Estimation}
The steering vectors obtained from local sensing and reflection processing are used to estimate the spatial frequencies along the corresponding array axes. For the sparse coprime HRIS geometry, the difference coarray contains missing samples and may lead to rank-deficient covariance matrices. Therefore, coarray-domain spatial smoothing is applied to restore the covariance rank and enable subspace-based angle estimation with Root-MUSIC \cite{ref5,ref4}.

After spatial smoothing, Root-MUSIC identifies the signal roots closest to the unit circle. The phase of each selected root provides an estimate of the corresponding spatial frequency. For the BS and UE ULAs, the estimated AoA and AoD are obtained as
\begin{equation}
    \hat{\theta}_\text{BS}=\arcsin\!\left(\frac{\hat{\mu}_\text{BS}}{\pi}\right),
    \quad \hat{\phi}_\text{UE}=\arcsin\!\left(\frac{\hat{\psi}_\text{UE}}{\pi}\right),
\end{equation}
where $\hat{\mu}_\text{BS}=\angle z_\text{BS}$ and $\hat{\psi}_\text{UE}=\angle z_\text{UE}$ are obtained from the corresponding Root-MUSIC roots.

For the HRIS URA, the elevation angles are first estimated from the vertical spatial frequencies as
\begin{equation}
    \hat{\theta}_{\text{ris}_D}=\arccos\!\left(\frac{\hat{\psi}_{\text{ris}_D}}{\pi}\right),
    \quad
    \hat{\theta}_{\text{ris}_A}=\arccos\!\left(\frac{\hat{\psi}_{\text{ris}_A}}{\pi}\right).
\end{equation}

Then, using $\mu_{\text{ris}}=\pi\sin(\theta_{\text{ris}})\sin(\phi_{\text{ris}})$, the azimuth angles are estimated as
\begin{equation}
\begin{split}
    \hat{\phi}_{\text{ris}_D}
    &=\arcsin\!\left(
    \frac{\hat{\mu}_{\text{ris}_D}}{\pi\sin(\hat{\theta}_{\text{ris}_D})}
    \right), \\
    \hat{\phi}_{\text{ris}_A}
    &=\arcsin\!\left(
    \frac{\hat{\mu}_{\text{ris}_A}}{\pi\sin(\hat{\theta}_{\text{ris}_A})}
    \right),
\end{split}
\end{equation}
where $\hat{\mu}_{\text{ris}_D}=\angle z_{y,D}$, $\hat{\mu}_{\text{ris}_A}=\angle z_{y,A}$, $\hat{\psi}_{\text{ris}_D}=\angle z_{z,D}$, and $\hat{\psi}_{\text{ris}_A}=\angle z_{z,A}$.

A step-by-step summary of the proposed HRIS parameter estimation and data fusion framework is provided in Algorithm \ref{alg:proposed_hris}.

\section{Simulation Results}

\begin{figure}[t]
    \centering
    \begin{minipage}{0.5\linewidth}
        \centering
        \includegraphics[width=\linewidth]{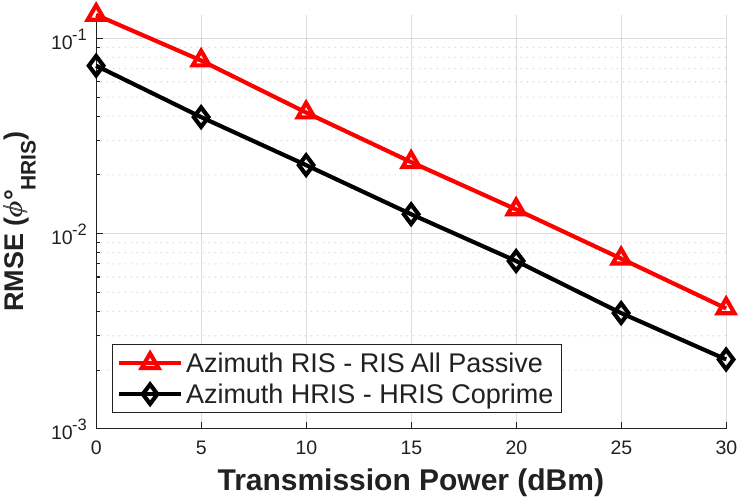}
        \caption{RMSE of the estimated arrival azimuth angle in the RIS.}
        \label{fig:aziRIS}
    \end{minipage}\hfill
    \begin{minipage}{0.5\linewidth}
        \centering
        \includegraphics[width=\linewidth]{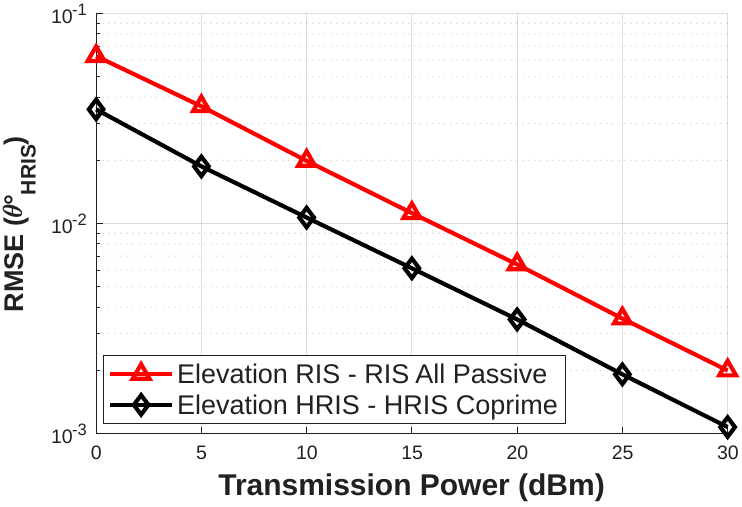}
        \caption{RMSE of the estimated arrival elevation angle in the RIS.}
        \label{fig:elevRIS}
    \end{minipage}
\end{figure}

\begin{figure}[t]
    \centering
    \begin{minipage}{0.5\linewidth}
        \centering
        \includegraphics[width=\linewidth]{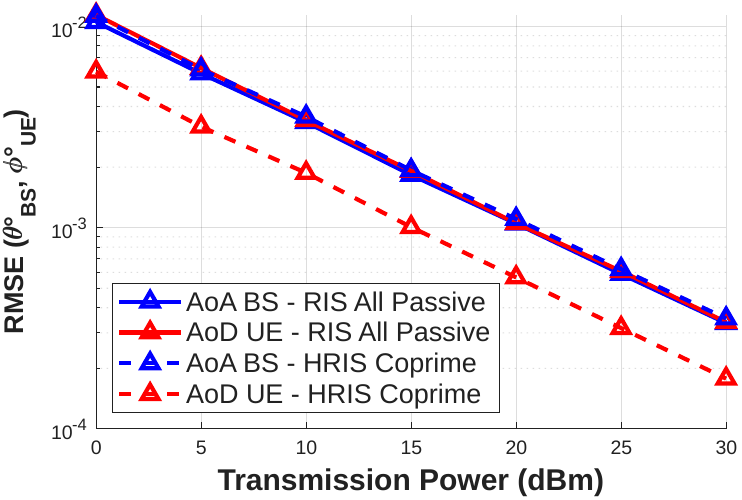}
        \caption{RMSE of the estimated departure UE and arrival BS angles.}
        \label{fig:terminals}
    \end{minipage}\hfill
    \begin{minipage}{0.5\linewidth}
        \centering
        \includegraphics[width=\linewidth]{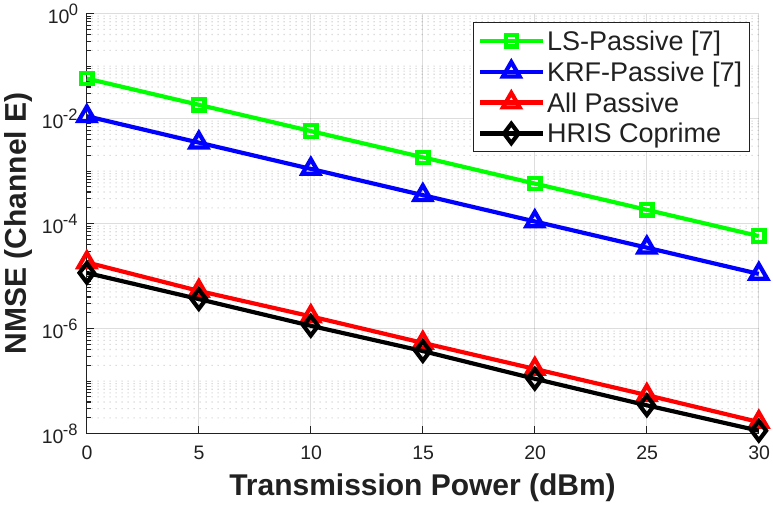}
        \caption{NMSE of the estimated channel $\hat{\bm{E}}$.}
        \label{fig:NMSE}
    \end{minipage}
\end{figure}

In this section, we evaluate the performance of the proposed parameter estimation scheme through Monte Carlo simulations. The proposed coprime HRIS architecture is compared with a fully passive RIS baseline. In the passive baseline, all RIS elements are used only for reflection and no local sensing is available; therefore, all parameters must be inferred from the cascaded signal observed at the BS. In contrast, the proposed HRIS activates only a small subset of elements arranged according to a sparse L-shaped coprime geometry, while the remaining elements continue to operate as passive reflectors. This design provides two important advantages. First, the active elements locally sense the UE-HRIS link before the signal experiences the second-hop path loss. Second, because only a few selected elements require RF chains and sensing circuitry, the architecture avoids the high hardware complexity of fully connected or subarray-based hybrid surfaces.

The simulation parameters for the uplink MIMO scenario are summarized in Table \ref{tab:sim_params}. The distance-dependent path loss (PL) model is given by $\text{PL}=\beta_0(d/d_0)^{-\alpha}$, where $\beta_0$ is the reference path loss at distance $d_0$ and $\alpha$ is the path loss exponent. The large-scale fading is incorporated into the channel matrices as $\bm{H} \leftarrow \sqrt{\text{PL}_\text{RB}}\bm{H}$ and $\bm{G} \leftarrow \sqrt{\text{PL}_\text{UR}}\bm{G}$, where $\text{PL}_\text{RB}$ and $\text{PL}_\text{UR}$ denote the RIS-BS and UE-RIS path losses, respectively \cite{ref6}.

\begin{table}[htbp]
\centering
\caption{Simulation Parameters}
\label{tab:sim_params}
\begin{tabular}{lcc}
\hline
\textbf{Parameter} & \textbf{Symbol} & \textbf{Value} \\
\hline
BS antennas (ULA) & $M$ & $10$ \\
UE antennas (ULA) & $Q$ & $10$ \\
Passive RIS elements (URA) & $N_y \times N_z$ & $11 \times 11$ \\
Active sensing elements (L-shaped) & $N_A$ & $11$ \\
Power splitting factor & $\rho$ & $0.5$ \\
UE-RIS distance & $d_\text{UR}$ & $100$ m \\
RIS-BS distance & $d_\text{RB}$ & $100$ m \\
Path loss exponent & $\alpha$ & $2.2$ \\
Reference path loss at $d_0 = 1$ m & $\beta_0$ & $-20$ dB \\
Receiver noise & -- & $-100$ dBm \\
\hline
\end{tabular}
\end{table}

The performance metric is the root mean square error (RMSE) of the estimated angular parameters, averaged over $2000$ independent Monte Carlo runs. Unless otherwise stated, the UE transmit power $P_t$ is varied from $0$ dBm to $30$ dBm. For the UE-HRIS link, both local sensing and reflection processing provide estimates of the same angular parameters. Therefore, the final estimates are obtained by averaging the sensing-based and reflection-based estimates as
\begin{equation}
\begin{split}
    \hat{\phi}_\text{UE}
    &= \frac{\hat{\phi}_\text{UE}^{\mathbb{S}}+\hat{\phi}_\text{UE}^{\text{Ref}}}{2}, \quad
    \hat{\phi}_{\text{ris}_A}
    = \frac{\hat{\phi}_{\text{ris}_A}^{\mathbb{S}}+\hat{\phi}_{\text{ris}_A}^{\text{Ref}}}{2}, \\
    \hat{\theta}_{\text{ris}_A}
    &= \frac{\hat{\theta}_{\text{ris}_A}^{\mathbb{S}}+\hat{\theta}_{\text{ris}_A}^{\text{Ref}}}{2}.
\end{split}
\end{equation}

This simple fusion rule exploits the complementary nature of the two observation mechanisms: the sensing branch benefits from a stronger single-hop UE-HRIS observation, whereas the reflection branch captures the effect of the complete RIS aperture through the cascaded channel.

Figures \ref{fig:aziRIS}--\ref{fig:terminals} report the angular RMSE performance. The proposed HRIS with $N_A=11$ active coprime sensing elements consistently improves the estimation of the UE-HRIS parameters, namely $\phi_\text{UE}$, $\phi_{\text{ris}_A}$, and $\theta_{\text{ris}_A}$, compared with the fully passive RIS. The improvement is mainly due to the local sensing capability of the active elements. In the passive architecture, the UE-HRIS information is available only after propagation through the complete UE-RIS-BS cascaded channel, which suffers from double path loss and stronger noise amplification. In the proposed HRIS, the active elements observe the incident UE signal directly at the surface, resulting in a higher effective sensing SNR for the UE-HRIS angular parameters.

The BS AoA $\theta_\text{BS}$ shows a different behavior because it is associated with the HRIS-BS link and is mainly inferred from the signal arriving at the BS. Consequently, both the passive RIS and HRIS schemes rely on the reflected/cascaded observation for this parameter, leading to similar BS-side angular performance. This result is expected and confirms that the main advantage of the proposed HRIS is not obtained by replacing the BS observation, but by complementing it with local sensing at the HRIS.

From a hardware perspective, the proposed coprime HRIS provides this performance gain with only a small number of active elements. For the $11\times 11$ RIS considered in Table \ref{tab:sim_params}, the surface has $N=121$ reflecting elements, but only $N_A=11$ are connected to RF chains. Thus, less than ten percent of the elements require active sensing hardware. This is significantly simpler than fully connected HRIS designs, where every element is connected to sensing or RF circuitry, and also simpler than subarray-based architectures, where multiple groups of elements require dedicated combiners, switches, or RF chains. The sparse coprime arrangement preserves a large virtual aperture for angular estimation while keeping the number of active components small. As a result, the proposed design achieves a favorable trade-off among estimation accuracy, reflection aperture, power consumption, and implementation cost.

The impact of the improved angular estimates on cascaded channel reconstruction is shown in Fig. \ref{fig:NMSE}, where the normalized mean square error (NMSE) of the structured matrix $\bm{E}=\bm{G}^T\diamond\bm{H}$ is evaluated. The proposed HRIS coprime scheme is compared with the fully passive RIS and with conventional unstructured matrix estimators, including direct least squares (LS) and Khatri-Rao factorization (KRF) \cite{refLSKRF}. The proposed method achieves the lowest NMSE across the considered transmit-power range. This gain occurs because the angular parameters estimated through Root-MUSIC are used to reconstruct the channels according to their physical Kronecker steering structure, instead of treating the cascaded channel as an arbitrary unstructured matrix. Therefore, the proposed HRIS not only improves angular estimation but also enhances the reconstruction of the cascaded channel.

\subsection{Impact of the Power Splitting Parameter}

\begin{figure}[t]
    \centering
    \begin{minipage}{0.5\linewidth}
        \centering
        \includegraphics[width=\linewidth]{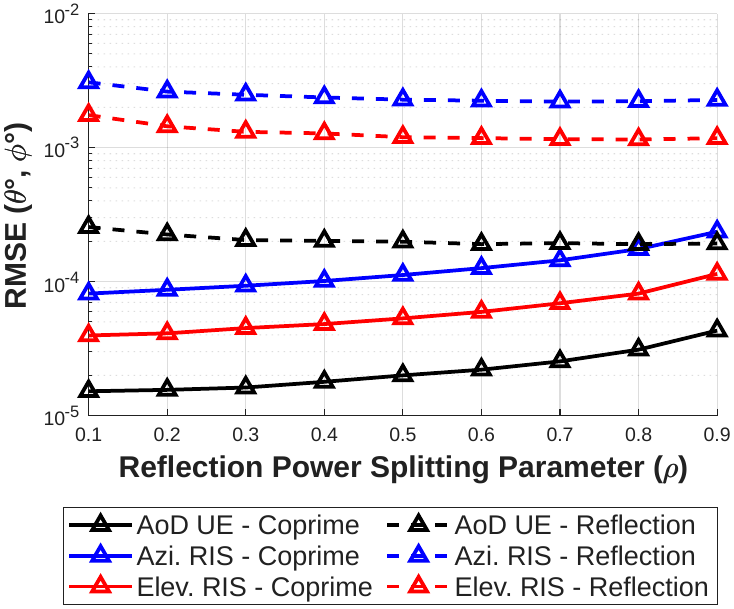}
        \caption{Impact of the power splitting parameter $\rho$ on local sensing and reflection-based angular estimation.}
        \label{fig:rhoChange}
    \end{minipage}\hfill
    \begin{minipage}{0.5\linewidth}
        \centering
        \includegraphics[width=\linewidth]{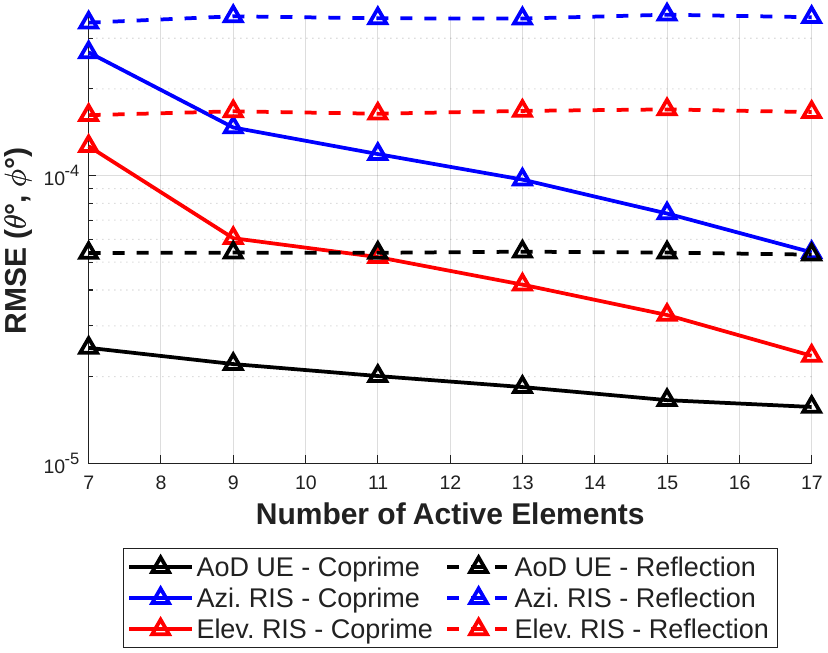}
        \caption{Impact of the number of active elements in the HRIS.}
        \label{fig:elementsChange}
    \end{minipage}
\end{figure}

Next, we evaluate the effect of the power splitting parameter $\rho$ on the estimation accuracy. The system configuration follows Table \ref{tab:sim_params}, while the UE transmit power is fixed at $P_t=15$ dBm. The parameter $\rho$ is varied from $0.1$ to $0.9$. For an active HRIS element, $\rho$ controls the fraction of incident power reflected toward the BS, whereas $1-\rho$ controls the fraction absorbed for local sensing. Therefore, $\rho$ directly determines the trade-off between reflection quality and sensing quality.

Figure \ref{fig:rhoChange} compares the RMSE of the angular parameters estimated from local sensing at the HRIS and from the reflected cascaded channel at the BS. The solid curves correspond to local sensing estimates, while the dashed curves correspond to reflection-based estimates. As $\rho$ increases, more signal power is reflected toward the BS, improving the cascaded-channel SNR and reducing the RMSE of the reflection-based estimates. Conversely, increasing $\rho$ leaves less power for the active sensors, which reduces the local sensing SNR and degrades the sensing-based estimates.

This behavior highlights an important design trade-off in HRIS systems. A small value of $\rho$ favors sensing but weakens the reflected link, whereas a large value of $\rho$ favors reflection but weakens local sensing. The proposed coprime HRIS remains attractive in this regime because it can exploit both sources of information. Even when one branch becomes less accurate due to power splitting, the other branch can provide complementary estimates, improving robustness compared with a fully passive architecture that relies only on the reflected signal.

\subsection{Impact of the Number of Active Elements}

Finally, we analyze the effect of the number of active sensing elements. In this experiment, the general simulation parameters are kept as in Table \ref{tab:sim_params}, with $\rho=0.5$. The HRIS is modeled as a larger URA with $N_y=21$ and $N_z=21$, corresponding to $N=441$ reflecting elements. The number of active elements is varied from $N_A=7$ to $N_A=17$ by expanding the sparse coprime geometry.

Figure \ref{fig:elementsChange} shows that the local sensing RMSE decreases as the number of active elements increases. This improvement is expected because adding active sensors enlarges the effective coprime aperture and increases the number of available spatial samples. Consequently, the spatial smoothing and Root-MUSIC stages obtain more reliable covariance information and achieve better angular resolution. Importantly, this improvement is obtained without activating the entire HRIS panel.

On the other hand, the reflection-based curves remain nearly constant as $N_A$ changes. This occurs because the reflected signal still uses almost the entire HRIS aperture. Since the active subset is very small compared with the total number of reflecting elements, increasing $N_A$ from 7 to 17 has a negligible impact on the overall reflection gain for a surface with $N=441$ elements. This confirms that the proposed architecture can improve sensing accuracy by adding only a few active elements while preserving the passive beamforming capability of the HRIS.

Overall, the simulation results demonstrate that the proposed coprime HRIS achieves improved estimation performance with substantially reduced hardware complexity. The sparse active array provides local sensing and large virtual-aperture processing, while the passive elements maintain the reflection gain. This combination makes the architecture more practical than fully active or densely connected hybrid surfaces and more accurate than fully passive RIS designs for channel parameter estimation.

\section{Conclusions}
This paper proposed a two-dimensional channel parameter estimation scheme for HRIS-assisted MIMO systems using a sparse coprime array under a dominant LoS model. By leveraging matrix factorizations and spatial smoothing, the cascaded channel was decoupled and rank-deficiencies resolved. Simulations demonstrated that this architecture outperforms fully passive baselines, offering robust estimation that improves sensing resolution without compromising reflection gains. To bridge the gap toward practical deployments, future work can extend this framework to rich-scattering non-line-of-sight (NLoS) conditions and dynamic environments, accounting for user mobility and Doppler spreads. Additionally, derive analytical methods for dynamic power-splitting optimization and expand the architecture to encompass multi-user, multi-HRIS scenarios with more practical, asymmetric antenna configurations.




\end{document}